\documentclass[a4paper,twoside,12pt]{article}
\input{obsjkt.sty}

\newcommand{\Msun}{\ensuremath{~{\rm M}_\odot}}                   
\newcommand{\Rsun}{\ensuremath{~{\rm R}_\odot}}                   
\newcommand{\rhosun}{\ensuremath{~\rho_\odot}}                    
\newcommand{\Teff}{\ensuremath{T_{\rm eff}}}                      
\newcommand{\FeH}{\ensuremath{\rm [Fe/H]}}                        
\newcommand{\logg}{\ensuremath{\log g}}                           
\newcommand{\degr}{\ensuremath{^\circ}}                           
\renewcommand{\kms}{~km~s$^{-1}$}                                 
\renewcommand{\cd}{~d$^{-1}$}                                     
\newcommand{\etal}{\textit{et al.}}                               

\newcommand{\tess}{\textit{TESS}}

\newcommand{\gaia}{\textit{Gaia}}

\newcommand{\Msunnom}{\hbox{$\mathcal{M}^{\rm N}_\odot$}}
\newcommand{\Rsunnom}{\hbox{$\mathcal{R}^{\rm N}_\odot$}}
\newcommand{\Lsunnom}{\hbox{$\mathcal{L}^{\rm N}_\odot$}}

\usepackage{rotating}

\begin{document} 

\OBSheader{Rediscussion of eclipsing binaries: RR Lyn}{J.\ Southworth}{2021 Dec}

\OBStitle{Rediscussion of eclipsing binaries. Paper VII. \\ Delta Scuti, Gamma Doradus and tidally-perturbed pulsations in RR Lyncis}

\OBSauth{John Southworth}

\OBSinstone{Astrophysics Group, Keele University, Staffordshire, ST5 5BG, UK}


\OBSabstract{RR~Lyn is a detached eclipsing binary with a 9.95~d orbit containing two A-stars: one metallic-lined and one possibly metal-poor. We use the light curve from the \tess\ satellite and two sets of published radial velocity measurements to determine the properties of the system to high precision. We find masses of $1.939 \pm 0.007$ and $1.510 \pm 0.003$\Msun, and radii of $2.564 \pm 0.019$ and $1.613 \pm 0.013$\Rsun. After adjusting published effective temperatures upwards by 200~K we find a good agreement with theoretical models for a solar chemical composition and an age of 1~Gyr, and a distance slightly shorter than expected from the \gaia\ EDR3 parallax. The light curve of RR~Lyn shows clear evidence for pulsations. We measure 35 pulsation frequencies and attribute the higher frequencies to $\delta$\,Scuti-type pulsations, and the intermediate frequencies to $\gamma$\,Doradus-type pulsations (some of which may be tidally perturbed). The lower frequencies may be tidally excited pulsations in RR\,Lyn or alternatively of instrumental origin. Most or all of these pulsations are likely to arise in the secondary star. RR~Lyn is one of the few eclipsing binaries known to have well-established properties and to exhibit multiple types of pulsations.}


\section*{Introduction}

Eclipsing binary stars are our primary source of direct measurements of the masses and radii of normal stars \cite{Andersen++90apj,Torres++10aarv}. Detached eclipsing binaries (dEBs) are of value as their properties can be compared to the predictions of theoretical models of stellar evolution in order to guide the refinement of these models \cite{ClaretTorres18apj,Tkachenko+20aa}.

Another type of object well suited to probing the physical properties of stars, in particular their interior structure, is the pulsating star \cite{Aerts++10book}. Detected oscillation frequencies in these objects may be compared to theoretical models to constrain properties such as their densities, ages and rotational profiles \cite{Aerts+03sci,Briquet+07mn,Garcia+13aa,Bedding+20nat}.

An obvious goal is to combine these two types of analysis by studying dEBs containing pulsating stars, in order to wield as many constraints on stellar theory as possible. This has now been achieved for many types of pulsator including $\delta$\,Scuti stars \cite{Me+11mn,Hambleton+13mn,Maceroni+14aa,Guo+16apj}, $\gamma$\,Doradus stars \cite{Maceroni+13aa,Guo+19apj}, slowly-pulsating B-stars \cite{Clausen96aa}, $\beta$\,Cephei pulsators \cite{Me+20mn,LeeHong21aj,Me++21mn} and red giants with solar-like oscillations \cite{Gaulme+16apj,Themessl+18mn,Benbakoura+21aa}. The binarity of these systems may also lead to tidal perturbation or excitation of their oscillations \cite{Maceroni+09aa,Fuller17mn,Bowman+19apj,Fuller+20mn}.

The pulsation type most commonly detected in stars in dEBs is $\delta$\,Scuti \cite{CampbellWright00apj,Baglin+73aa,Breger00aspc}. These are main-sequence or subgiant stars with masses of 1.5 to 2.5\Msun\ and effective temperature (\Teff) values of 7100 to 9000\,K \cite{Murphy+19mn}. They show low-amplitude radial and non-radial pressure-mode oscillations with periods of 0.015 to 0.33\,d \cite{Breger00aspc,Grigahcene+10apj} that can be used to determine their density \cite{Garcia+13aa,Bedding+20nat,Garcia+15apj}.

Another class of oscillation that can occur in late-A and early-F stars is $\gamma$\,Doradus pulsations \cite{Kaye+99pasp}. These are gravity-mode pulsations that are sensitive to the interior properties of the stars \citep{Aerts++10book}. The pulsation periods range from 0.3\,d to 4\,d and the amplitudes are up to 0.1\,mag \citep{Grigahcene+10apj,Henry++07aj}. $\delta$\,Scuti and $\gamma$\,Dor oscillations can coexist \cite{Grigahcene+10apj} and such stars are called hybrid pulsators. Balona \etal\ \cite{Balona++15mn} found that all $\delta$\,Scuti stars show low-frequency oscillations in high-quality data, so hybrid pulsation may be the standard situation.

In this work we present the detection of $\delta$\,Scuti and $\gamma$\,Dor pulsations in the dEB RR~Lyn. This is part of our work to systematically reanalyse dEBs in the DEBCat\footnote{\texttt{https://www.astro.keele.ac.uk/jkt/debcat/}} catalogue \cite{Me15debcat} (see Paper I of the series \cite{Me20obs}).


\begin{table}[t]
\caption{\em Basic information on RR~Lyn \label{tab:info}}
\centering
\begin{tabular}{lll}
{\em Property}                 & {\em Value}            & {\em Reference}                \\[3pt]
Bright Star Catalogue          & HR 2291                & \cite{HoffleitJaschek91}       \\
Henry Draper designation       & HD 44691               & \cite{CannonPickering18anhar2} \\
\textit{Gaia} EDR3 designation & 997809280404484480     & \cite{Gaia21aa}                \\
\textit{Gaia} EDR3 parallax    & $12.416 \pm 0.092$ mas & \cite{Gaia21aa}                \\
\tess\ designation             & TIC 11491822           & \cite{Stassun+19aj}            \\
$B_T$ magnitude                & $5.790 \pm 0.014$      & \cite{Hog+00aa}                \\
$V_T$ magnitude                & $5.585 \pm 0.009$      & \cite{Hog+00aa}                \\
$J$ magnitude                  & $5.471 \pm 0.290$      & \cite{Cutri+03book}            \\
$H$ magnitude                  & $5.066 \pm 0.020$      & \cite{Cutri+03book}            \\
$K_s$ magnitude                & $4.993 \pm 0.016$      & \cite{Cutri+03book}            \\
Spectral type                  & A3/A7V/F2 + F0\,V      & \cite{LevatoAbt78pasp,Khaliullin++01arep}         \\[10pt]
\end{tabular}
\end{table}

\section*{RR Lyncis}

RR~Lyn is a bright dEB containing two stars of significantly different mass and radius in an orbit with a period of 9.95\,d and a small eccentricity. It was discovered to be a spectroscopic binary from observations in early 1911 collected by Adams \cite{Adams12apj}. A first period determination and single-lined spectroscopic orbit was given by Harper \cite{Harper15pdo}, under the moniker 1149 Groombridge, based on 30 photographic spectra. Another single-lined orbit was obtained by Douglas \& Popper \cite{DouglasPopper63pasp} and double-lined orbits have since been published by Popper \cite{Popper71apj}, Kondo \cite{Kondo76antok}, Tomkin \& Fekel \cite{TomkinFekel06aj} and Bensch \etal\ \cite{Bensch+14ibvs}. The last two papers are of particular interest as they present high-quality radial velocities (RVs), obtained with \'echelle spectrographs, that can be included in our analysis.

The discovery of eclipses in RR~Lyn was announced by Huffer \cite{Huffer31paas}, where it was named Boss 1607. Photoelectric light curves have subsequently been obtained by Magalashvili \& Kumsishvili \cite{MagalashviliKumsishvili59abaob}, Botsula \cite{Botsula60baoe}, Linnell \cite{Linnell66aj}, Lavrov \etal\ \cite{Lavrov++88trkaz} and Khaliullin \etal\ \cite{Khaliullin++01arep}. Those of Linnell \cite{Linnell66aj} were in the $UBV$ system and are tabulated in that work, so may be used in future to determine the individual $UBV$ magnitudes of the two stars.

The presence of a third body in the system was suggested by Khaliullin \& Khaliullina \cite{KhaliullinKhaliullina02arep} based on deviations of the eclipse times from a linear ephemeris. These authors suggested a period of $39.7 \pm 4.2$~yr, an extreme orbital eccentricity of $e = 0.96 \pm 0.02$, and a minimum mass of $0.10 \pm 0.02$\Msun. The putative tertiary component should imprint deviations of 0.002\,d on eclipse times \cite{KhaliullinKhaliullina02arep} and 2.5\kms\ on the systemic velocity \cite{TomkinFekel06aj} but the evidence for either is weak. The referee has instead found evidence for a light-time effect due to a third body on an orbit of roughly 65\,yr period; a detailed eclipse timing analysis of the system is warranted.

Another method for detecting third components is to search for third light ($\ell_3$) when analysing light curves of eclipses \cite{Me21obs4}. In the case of RR~Lyn third light has been found by Linnell \cite{Linnell66aj} and Budding \cite{Budding74apss} but was not needed in the analyses by Botsula \cite{Botsula68sovast} and Khaliullin \etal\ \cite{Khaliullin++01arep}.

RR~Lyn has been known for a long time \cite{Roman49apj,Cowley+69aj,AbtBidelman69apj} to also exhibit clear chemical peculiarities of the Am type \cite{TitusMorgan40apj,Conti70pasp} in its spectrum. Popper \cite{Popper71apj} classified it as A3 based on the calcium K line and F0 based on the hydrogen Balmer lines. Levato \& Abt \cite{LevatoAbt78pasp} noted its metallic-line nature and gave its spectral type as A3 based on the Ca\,II K line, A7\,V based on the Balmer lines and F2 based on the metal lines. Abt \& Morrell \cite{AbtMorrell95apjs} classified the system as A3/A8/A6. Khaliullin \etal\ \cite{Khaliullin++01arep} obtained spectral types photometrically using the $UBVR$ filter system, finding A6\,IV for the primary (hereafter star A) and F0\,V for the secondary (hereafter star B). They furthermore obtained \FeH\ values of $+0.31 \pm 0.08$ for star A and $-0.24 \pm 0.06$ for star B based on the manifestation of line blanketing effects in the $WBVR$ passbands.



\section*{Observational material}

\begin{figure}[t] \centering \includegraphics[width=\textwidth]{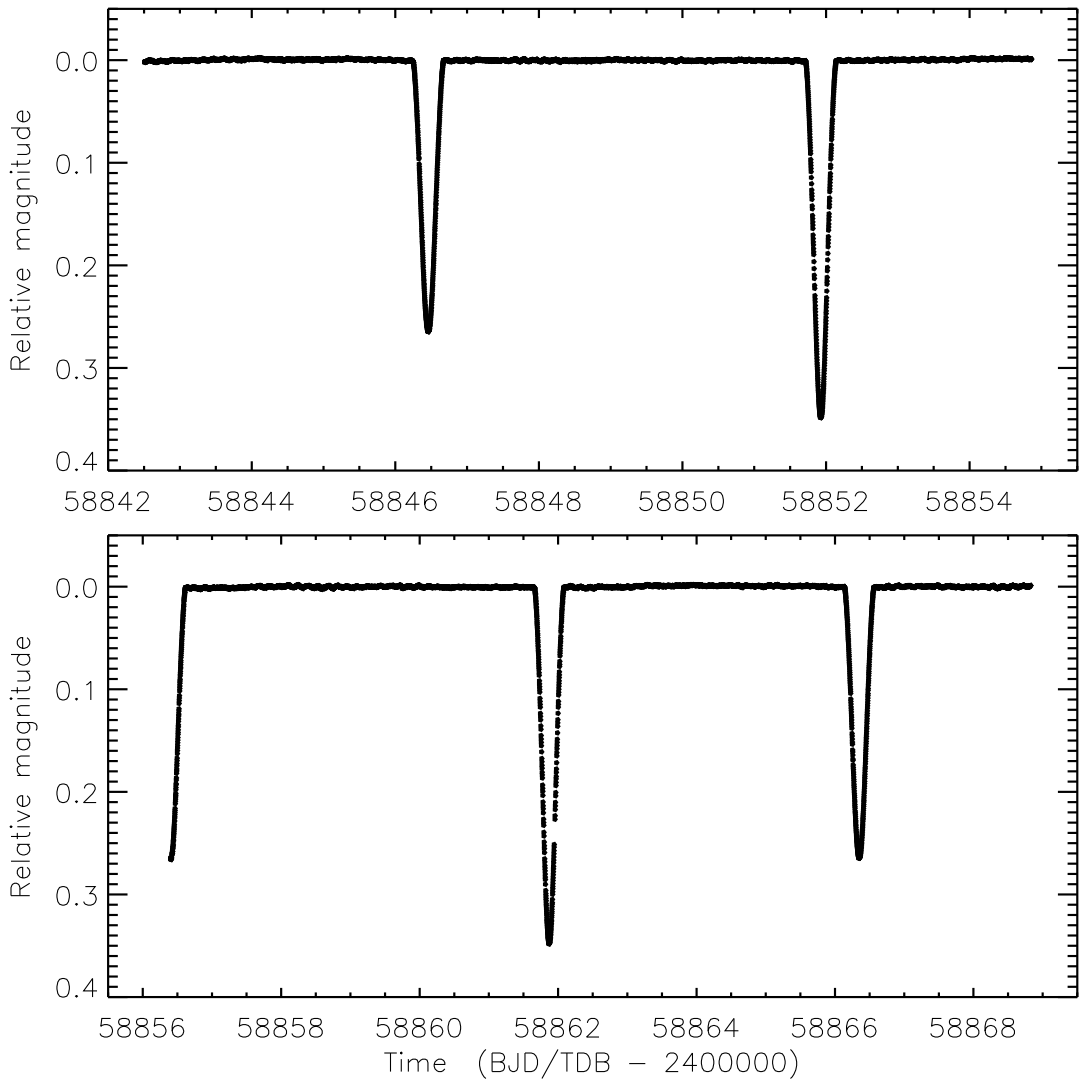} \\
\caption{\label{fig:time} \tess\ Sector 20 short-cadence SAP photometry of RR~Lyn.
The two panels show data from before and after the mid-sector pause.} \end{figure}


RR~Lyn was observed using camera 2 of the NASA \tess\ satellite \cite{Ricker+15jatis} in Sector 20, and no further observations are planned from this satellite. The light curve comprises 18\,954 datapoints obtained in short cadence mode \cite{Jenkins+16spie}, which were downloaded from the MAST archive\footnote{Mikulski Archive for Space Telescopes, \\ \texttt{https://mast.stsci.edu/portal/Mashup/Clients/Mast/Portal.html}} and converted to relative magnitude. All datapoints whose QUALITY flag was not zero were rejected, leaving 17\,552 observations.

As with previous papers of this series, we used the simple aperture photometry (SAP) version of the \tess\ data. This light curve contains two primary and two secondary eclipses observed in their entirety. One further secondary eclipse was only partially observed as it fell near the mid-sector pause for download of the data to Earth (Fig.\,\ref{fig:time}).


\section*{Analysis of the \tess\ light curve}

\begin{figure}[t] \centering \includegraphics[width=\textwidth]{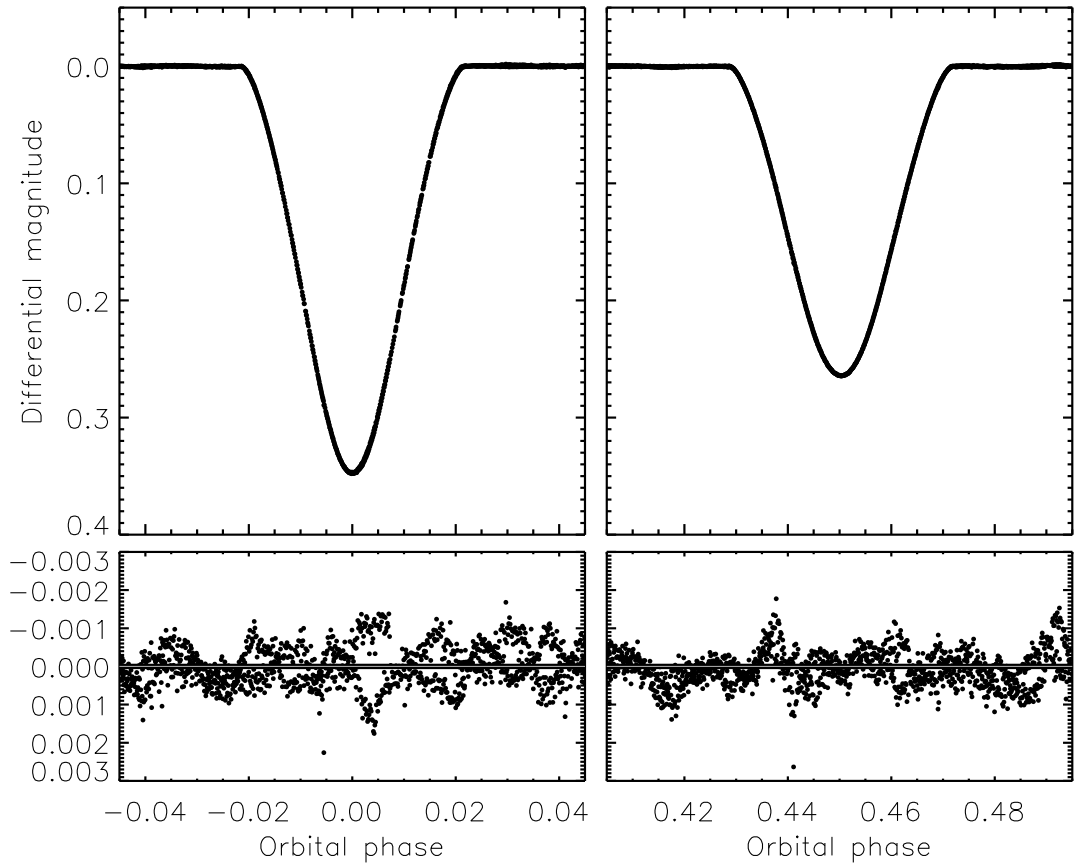} \\
\caption{\label{fig:tess} The \tess\ light curve of RR~Lyn (filled circles) around the primary (left) and
secondary (right) eclipses. The best fit is not plotted as it is indistinguishable from the data. The lower
panels show the residuals of the fit with the line of zero residual overplotted in white for clarity.} \end{figure}

The great majority of the data in the \tess\ light curve of RR~Lyn are far from an eclipse and contribute negligible constraints on the radii of the stars. We therefore cut from the light curve all datapoints more than 1.25\,d (approximately three times the eclipse duration) from the midpoint of the four eclipses that were fully observed. This left a total of 3553 datapoints for detailed analysis. We rescaled their errorbars to force a reduced $\chi^2$ of $\chi^2_\nu = 1$.

We then modelled the 3553 datapoints using version 41 of the {\sc jktebop}\footnote{\texttt{http://www.astro.keele.ac.uk/jkt/codes/jktebop.html}} code \cite{Me++04mn2,Me13aa}, which is appropriate for systems with well-separated stars \cite{Maxted+20mn}. By definition the primary eclipse is the deeper of the two, star~A is eclipsed during primary minimum, and star~B is eclipsed during secondary minimum. {\sc jktebop} is parameterised using the fractional radii of the stars ($r_{\rm A} = \frac{R_{\rm A}}{a}$ and $r_{\rm B} = \frac{R_{\rm B}}{a}$) where $R_{\rm A}$ and $R_{\rm B}$ are the true radii and $a$ is the semimajor axis of the relative orbit. We included their sum ($r_{\rm A}+r_{\rm B}$) and ratio ($k = {r_{\rm B}}/{r_{\rm A}}$) as fitted parameters, along with the orbital inclination ($i$), period ($P$), time of mid-primary-eclipse ($T_0$), and the central surface brightness ratio of the two stars ($J$). RR~Lyn shows a small but highly significant orbital eccentricity ($e$) which we accounted for by fitting for $e\cos\omega$ and $e\sin\omega$ where $\omega$ is the argument of periastron. Limb darkening (LD) was accounted for using the quadratic law \cite{Kopal50} with the linear coefficients of the two stars fitted and the quadratic coefficients fixed at theoretical values obtained from Claret \cite{Claret18aa}. We included third light ($\ell_3$) as a fitted parameter due to the possible presence of a tertiary star. The final fitted parameters were the coefficients of a straight line fit to the out-of-eclipse brightness of the system for each eclipse.

\begin{table} \centering
\caption{\em \label{tab:lc} Parameters of the best {\sc jktebop} fit to the \tess\ light curve of RR~Lyn, with
and without third light. The adopted solution is that including third light. The uncertainties are 1$\sigma$.
The primary eclipse time is given as BJD/TDB $-$ 2400000.}
\begin{tabular}{lr@{\,$\pm$\,}lr@{\,$\pm$\,}l}
{\em Parameter}                           & \multicolumn{2}{c}{\em Third light}           & \multicolumn{2}{c}{\em No third light}           \\[3pt]
{\it Fitted parameters:} \\
Time of primary eclipse                   &   58851.92622   & 0.00010                     &   58851.92622 & 0.00007                          \\
Orbital period (d)                        &      9.945120   & 0.000073                    &      9.945120 & 0.00066                          \\
Orbital inclination (\degr)               &      87.46      & 0.13                        &      87.18    & 0.07                             \\
Sum of the fractional radii               &       0.14206   & 0.00082                     &       0.14393 & 0.00071                          \\
Ratio of the radii                        &       0.6292    & 0.0069                      &       0.637   & 0.011                            \\
Central surface brightness ratio          &       0.816     & 0.032                       &       0.817   & 0.030                            \\
Third light                               &       0.036     & 0.023                       & \multicolumn{2}{c}{~~~~~~~~~~0.0 (fixed)}        \\
Linear LD coefficient star A              &       0.177     & 0.062                       &       0.240   & 0.029                            \\
Quadratic LD coefficient star A           & \multicolumn{2}{c}{~~~~~~~~~~0.25 (fixed)}    & \multicolumn{2}{c}{~~~~~~~~~~0.25 (fixed)}       \\
Linear LD coefficient star B              &       0.230     & 0.077                       &       0.312   & 0.096                            \\
Quadratic LD coefficient star B           & \multicolumn{2}{c}{~~~~~~~~~~0.22 (fixed)}    & \multicolumn{2}{c}{~~~~~~~~~~0.22 (fixed)}       \\
$e\cos\omega$                             &    $-$0.078061  & 0.000018                    &   $-$0.078043 & 0.000017                         \\
$e\sin\omega$                             &    $-$0.0016    & 0.0032                      &       0.0019  & 0.0034                           \\[5pt]
{\it Derived parameters:} \\
Fractional radius of star~A               &       0.08720   & 0.00063                     &       0.08790 & 0.00019                          \\
Fractional radius of star~B               &       0.05486   & 0.00045                     &       0.05603 & 0.00083                          \\
Orbital eccentricity                      &       0.078078  & 0.000076                    &      0.078067 & 0.000094                         \\
Argument of periastron (\degr)            &     178.8       & 2.3                         &         178.6 & 2.5                              \\
Light ratio                               &       0.3183    & 0.0060                      &        0.3247 & 0.0091                           \\
rms residual of the fit (mmag)            & \multicolumn{2}{c}{~~~~~~~~~~0.5137}          & \multicolumn{2}{c}{~~~~~~~~~~0.5158}             \\
\end{tabular}
\end{table}

The best fit is shown in Fig.\,\ref{fig:tess} and the measured parameters are given in Table\,\ref{tab:lc}. We include solutions calculated with $\ell_3$ fitted and with $\ell_3 = 0$ for reference. We adopt the solution with third light as it is a slightly better fit and is less affected by any imperfections in the sky background calculation during the reduction of the \tess\ data. The two solutions are consistent to within 1.4$\sigma$ for $r_{\rm B}$, which is the most discrepant parameter. Our adopted solution has a positive but insignificant $\ell_3$, so neither proves nor disproves the possible presence of a third body.

The uncertainties in the measured parameters were determined using Monte Carlo and residual-permutation algorithms \cite{Me++04mn,Me08mn}, and the larger of the two uncertainties for each parameter was retained. The pulsations were not explicitly accounted for in the {\sc jktebop} analysis so have the effect of contributing red noise to our results. In all cases the residual-permutation errorbars were significantly larger than the Monte Carlo errorbars, by factors of typically 4 to 6. This is likely due to the effects of the pulsations combined with having only a small number of eclipses observed by \tess. The net result is measurement of $r_{\rm A}$ to 0.7\% and $r_{\rm B}$ to 0.8\% precision. The measured fractional radii and their uncertainties are of similar size to previous values \cite{Khaliullin++01arep} but likely more reliable as they rest on data of greater number and much higher precision. The orbital phase of mid-secondary-eclipse is 0.4504. The residuals in Fig.\,\ref{fig:tess} are slightly larger during primary eclipse, which suggests that star B is the source of the pulsational brightness changes discussed below.



\section*{Analysis of published radial velocities}

\begin{figure}[t] \centering \includegraphics[width=\textwidth]{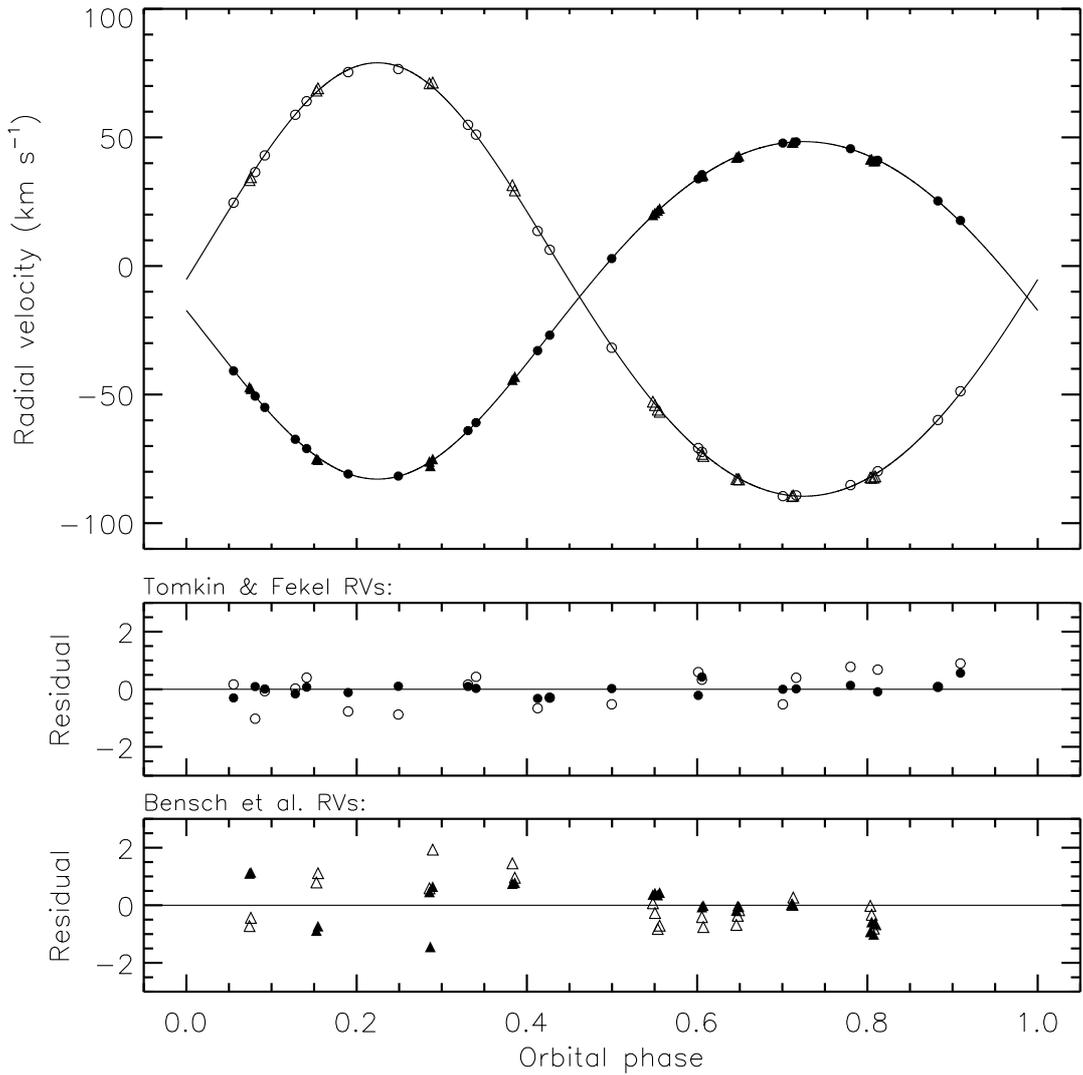} \\
\caption{\label{fig:rv} The spectroscopic orbit of RR~Lyn compared to the RVs from the two
different sources. Circles show RVs from Tomkin \& Fekel \cite{TomkinFekel06aj} and triangles
show RVs from Bensch \etal\ \cite{Bensch+14ibvs}. 2\kms\ has been subtracted from the RVs from
Bensch \etal\ \cite{Bensch+14ibvs} for display purposes, to place them on the same systemic
velocity as the RVs from Tomkin \& Fekel \cite{TomkinFekel06aj}. Filled symbols are for star~A
and open symbols for star~B. The solid lines show the fitted spectroscopic orbits for the stars.
The residuals are shown on an expanded scale, and separately for the two sources of RVs, in the
lower panels (labelled). Orbital phase zero is the time of primary eclipse.} \end{figure}

Two works have previously obtained and analysed RVs from high-dispersion spectra, and we have obtained these and fitted them with {\sc jktebop} to confirm and combine the results. Tomkin \& Fekel \cite{TomkinFekel06aj} presented 21 measurements for each component, neglecting a single zero-weight observation, with a scatter of 0.22\kms\ for star~A and 0.51\kms\ for star~B. Bensch \etal\ \cite{Bensch+14ibvs} obtained 37 spectra of which 23 had resolved lines for both components. Our own fit of these 23 pairs of RVs returned scatters of 0.70 and 0.78\kms\ for the two stars, respectively.

Under the presumption that it is best to combine datasets to obtain the most precise results, and bearing in mind that Bensch \etal\ \cite{Bensch+14ibvs} did not quote the velocity amplitudes from their fit to the RVs, we fitted the two datasets simultaneously with {\sc jktebop} (Fig.\,\ref{fig:rv}). We fitted for the velocity amplitudes ($K_{\rm A}$ and $K_{\rm B}$), $e\cos\omega$, $e\sin\omega$ and $T_0$. The uncertainties in the RVs were set to give $\chi^2_\nu = 1$ for each star in each dataset and the systemic velocities of the two stars in each dataset were fitted separately. We find $K_{\rm A} = 65.620 \pm 0.045$\kms\ and $K_{\rm B} = 84.28 \pm 0.13$\kms, plus values of $e\cos\omega$, $e\sin\omega$ and $T_0$ consistent with those from the previous section. The uncertainties were obtained using Monte Carlo simulations, although the formal errors of the fit are very similar (see Paper\,VI of this series \cite{Me21obs5}). The values of $K_{\rm A}$ and $K_{\rm B}$ are consistent with those from Tomkin \& Fekel \cite{TomkinFekel06aj} but have smaller errorbars.



\section*{Physical properties of RR Lyn}

\begin{table} \centering
\caption{\em Physical properties of RR~Lyn defined using the nominal solar
units given by IAU 2015 Resolution B3 (Ref.\ \cite{Prsa+16aj}). \label{tab:absdim}}
\begin{tabular}{lr@{\,$\pm$\,}lr@{\,$\pm$\,}l}
{\em Parameter}        & \multicolumn{2}{c}{\em Star A} & \multicolumn{2}{c}{\em Star B}    \\[3pt]
Mass ratio                                  & \multicolumn{4}{c}{$0.7790 \pm 0.0013$}       \\
Semimajor axis of relative orbit (\Rsunnom) & \multicolumn{4}{c}{$29.405 \pm 0.027$}        \\
Mass (\Msunnom)                             &  1.9394 & 0.0065      &  1.5100 & 0.0033      \\
Radius (\Rsunnom)                           &   2.564 & 0.019       &   1.613 & 0.013       \\
Surface gravity ($\log$[cgs])               &  3.9078 & 0.0063      &  4.2017 & 0.0071      \\
Density ($\!$\rhosun)                       &  0.1150 & 0.0025      &  0.3597 & 0.0089      \\
Synchronous rotational velocity ($\!\!$\kms)&  13.044 & 0.095       &   8.106 & 0.068       \\
Effective temperature (K)                   &    7770 & 200         &    7180 & 200         \\
Luminosity $\log(L/\Lsunnom)$               &   1.334 & 0.045       &   0.795 & 0.049       \\
$M_{\rm bol}$ (mag)                         &    1.40 & 0.11        &    2.75 & 0.12        \\
Distance (pc)                               & \multicolumn{4}{c}{$76.7 \pm 1.0$}            \\[5pt]
\end{tabular}
\end{table}

We have determined the physical properties of the system from the values of the quantities $r_{\rm A}$, $r_{\rm B}$, $i$, $e$, $P$, $K_{\rm A}$ and $K_{\rm B}$ measured above. This was done using standard formulae \cite{Hilditch01book} and the {\sc jktabsdim} code \cite{Me++05aa}, and resulted in the quantities shown in Table\,\ref{tab:absdim}. The masses of the stars are measured to precisions of 0.3\%, and their radii to 0.8\%.


A determination of the distance to the system and the luminosities of the stars needs their \Teff\ values. These were found to be ${\Teff}_{\rm(A)} = 7570 \pm 120$~K and ${\Teff}_{\rm(B)} = 6980 \pm 100$~K by Khaliullin \etal\ \cite{Khaliullin++01arep} using $WBVR$ photometry and a calibration of the \Teff\ scale \cite{Popper80araa}. To these we added the apparent magnitudes of the system given in Table\,\ref{tab:info}, and an interstellar extinction estimate of $E(B-V) = 0.002 \pm 0.002$\,mag obtained using the {\sc stilism}\footnote{\texttt{https://stilism.obspm.fr}} online tool (Lallement \etal\ \cite{Lallement+14aa,Lallement+18aa}). The resulting distance measurement is significantly shorter than that from the \gaia\ EDR3 parallax of the system ($80.54 \pm 0.59$~pc; Table\,\ref{tab:info}), and the \Teff\ values are low compared to theoretical predictions (see below). To (partially) alleviate these discrepancies we took the simple step of adding 200~K to the \Teff\ of both stars, and also increased the errorbars to $\pm$200~K. The adjusted \Teff\ of star~A is consistent with its spectral type of A7\,V \cite{PecautMamajek13apjs} and with a recent determination by Graczyk \etal\ \cite{Graczyk+19apj}. The surface brightness ratio found in the {\sc jktebop} analysis implies that the \Teff\ values of the stars differ by a smaller amount than this, $370 \pm 70$~K, suggesting that further work on this point is needed, preferably in the form of a detailed analysis using high-resolution spectroscopy.

\begin{figure}[t] \centering \includegraphics[width=\textwidth]{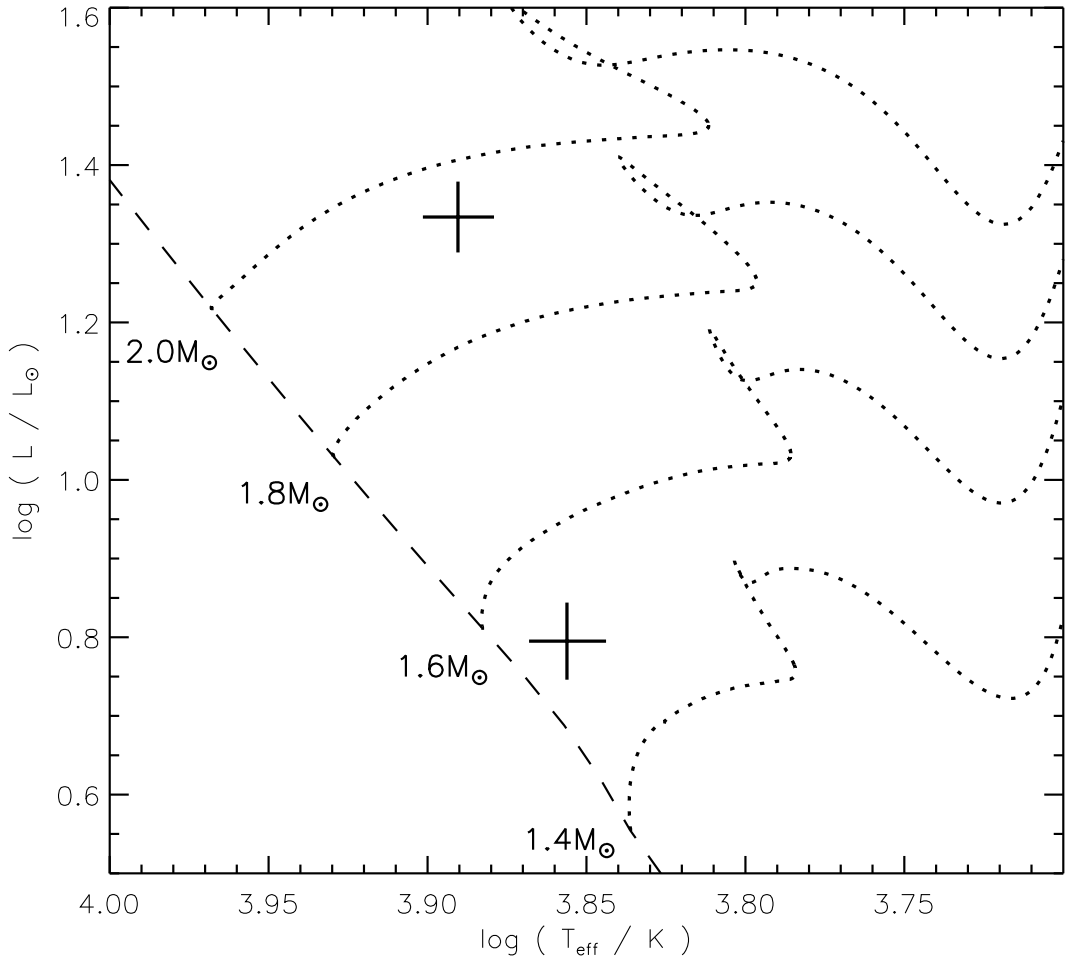} \\
\caption{\label{fig:hrd} Hertzsprung-Russell diagram showing the components of RR~Lyn
(solid crosses) and selected predictions from the PARSEC models \cite{Bressan+12mn}
(dotted lines) beginning at the zero-age main sequence. Models for 1.4, 1.6, 1.8 and
2.0\Msun\ are shown (labelled), all for a solar chemical composition.} \end{figure}

In order to gain a theoretical perspective of RR~Lyn, the masses, radii and \Teff\ values of the stars were compared to the predictions of the {\sc parsec} models \cite{Bressan+12mn} assuming a solar chemical composition. The models match the observed masses and radii for this composition and an age of $950 \pm 20$~Myr, where the errorbar is a fitting uncertainty which does not take into account any imperfections in the models. The higher \Teff\ values given in Table\,\ref{tab:info} agree very well with the theoretical predictions for this age. We also performed a comparison on the Hertzsprung-Russell diagram (Fig.\,\ref{fig:hrd}), finding that star~B is close to the zero-age main sequence but that star~A has evolved roughly half-way to the terminal-age main sequence.


\section*{Pulsation analysis}

\begin{figure}[t] \centering \includegraphics[width=\textwidth]{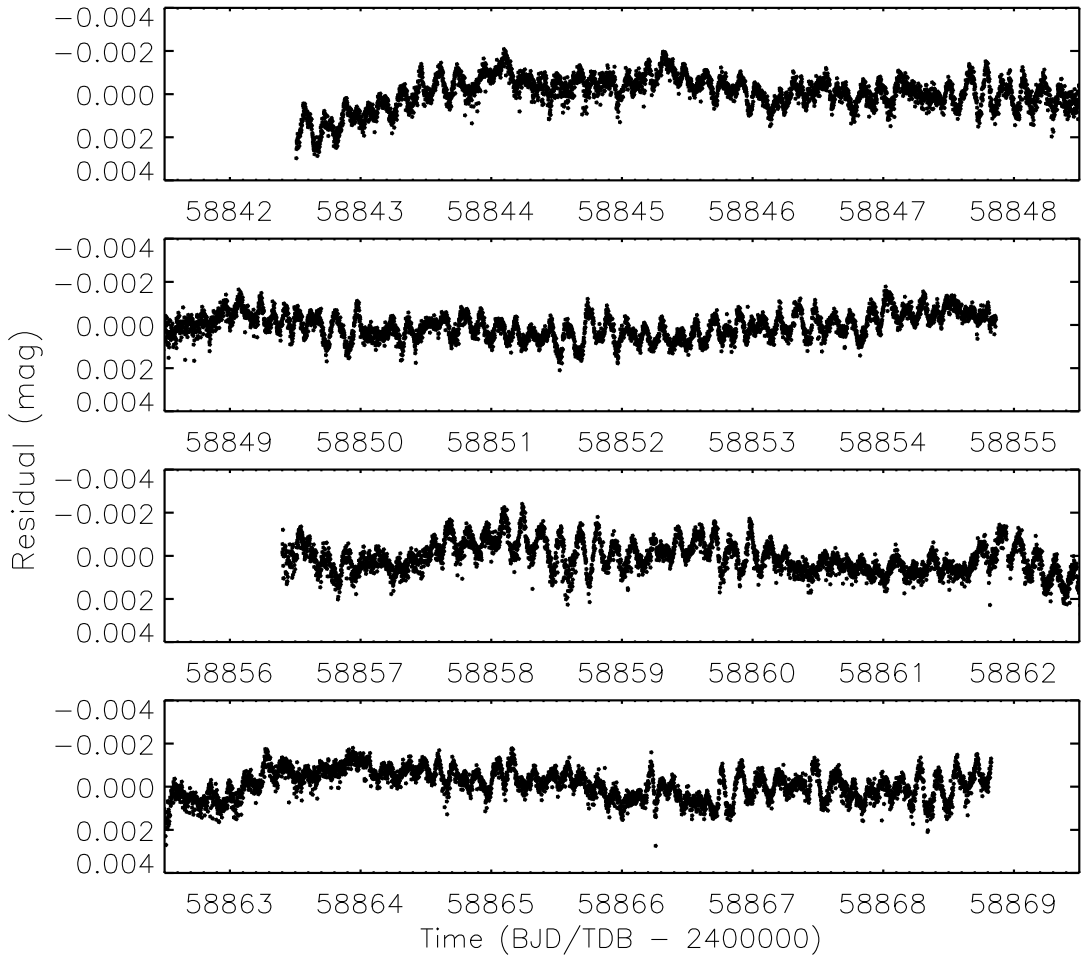} \\
\caption{\label{fig:resid} Residuals of the best fit to the full TESS light curve
of RR~Lyn, plotted so as to make the pulsations clear.} \end{figure}

The light curve in Fig.\,\ref{fig:time} shows evidence for short-period variability, which can be investigated using the residuals from the {\sc jktebop} fit. We therefore performed a fit to the full TESS data using {\sc jktebop} in order to remove the effects of binarity from the light curve. The residuals of this fit are shown in Fig.\,\ref{fig:resid}, where the details have been brought out by stretching the time axis over several panels. A short-period variation is obvious, and several longer-period variations are also present in these data.

To measure pulsation frequencies from the residuals we used version 1.2.0 of the {\sc period04} code \cite{LenzBreger05coast} to calculate a frequency spectrum from 0 to the Nyquist frequency of 360\cd. No significant periodicities above 19\cd\ were detected. We then selected significant frequencies in the spectrum and fitted sinusoids simultaneously to all of them. We included only those frequencies for which the signal-to-noise (S/N) is more than 10. This is much higher than the widely-used criterion of S/N $>$ 4 (Refs.\ \cite{Breger+93aa},\cite{Kuschnig+97aa}) but is sufficient to illustrate the general nature of the star. For reference the orbital frequency is 0.1006\cd\ and the Loumos \& Deeming \cite{LoumosDeeming78apss} frequency resolution is $2.5\,/\,\Delta T = 0.095$\cd\ where $\Delta T$ is the time interval covered by the data.


\begin{sidewaysfigure} \centering
\includegraphics[width=\textwidth]{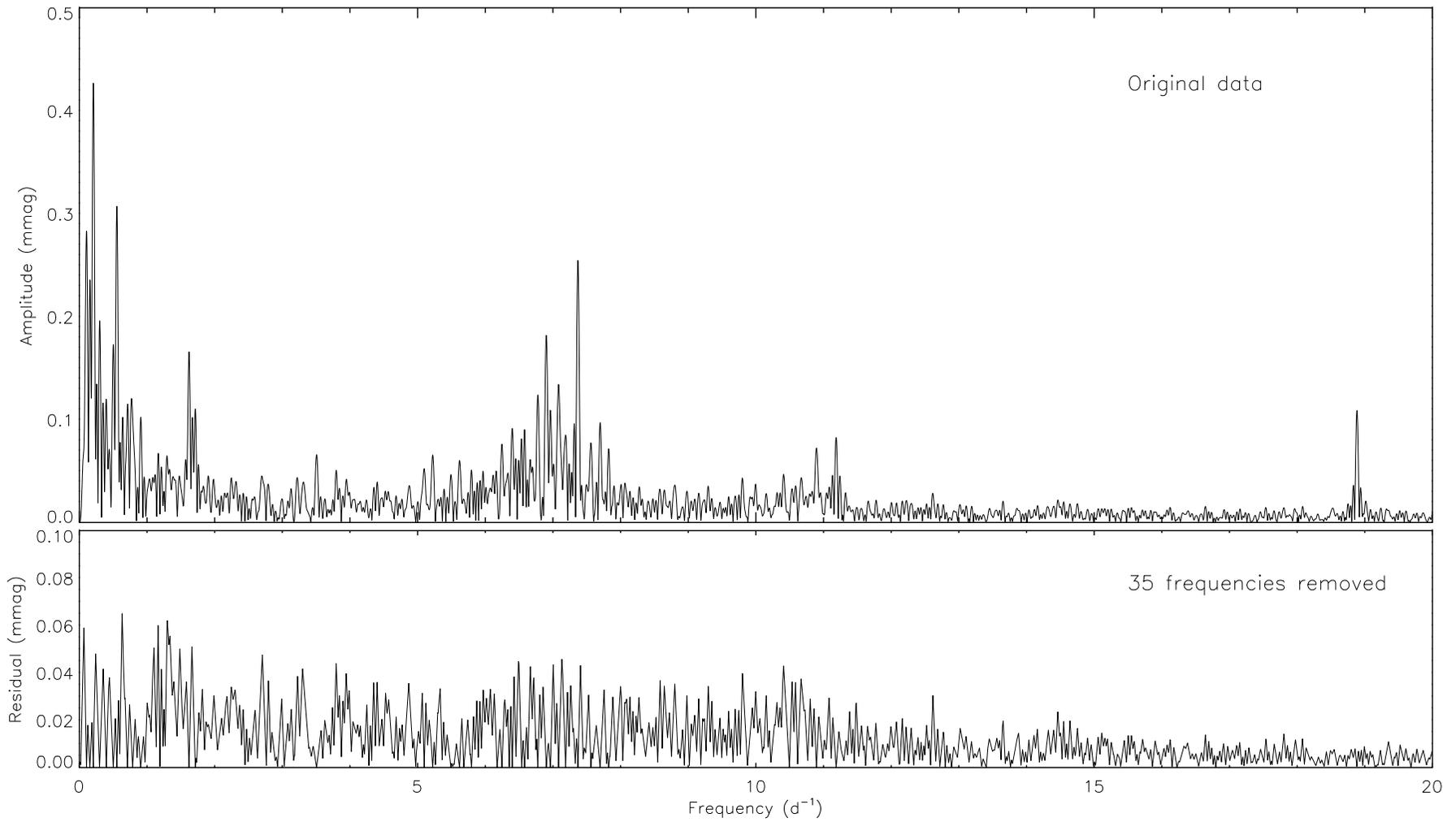}
\caption{\label{fig:freq} Amplitude spectrum of the \tess\ light curve of RR~Lyn.
Top: spectrum of the data after subtraction of the binary model.
Bottom: spectrum after subtraction of the binary model and the 35 frequencies
measured in this work.} \end{sidewaysfigure}


We measured a total of 35 frequencies from the data and calculated the amplitude and phase of each one (Table\,\ref{tab:freq}). Some of these are close to multiples of the orbital frequency ($f_{\rm orb}$) and are labelled in Table\,\ref{tab:freq}. The detected frequencies fall into three categories. Frequency spectra of the data before and after subtraction of the 35 frequencies are shown in Fig.\,\ref{fig:freq}.

The lowest three frequencies ($f_{ 1}$, $f_{ 2}$, $f_{ 3}$) are close to $f_{\rm orb}$, 1.5$f_{\rm orb}$ and 2$f_{\rm orb}$. These either arise from imperfections in the light curve model or the normalisation of the \tess\ data, or are pulsations induced by the orbital motion of the system \cite{Maceroni+09aa,Fuller17mn}. Because these frequencies are similar to the length of the time intervals over which RR\,Lyn was continuously monitored by \tess, we cannot be sure they arise from the target. More detailed analysis is necessary before claiming the reality of these signals.

Frequencies $f_4$ to $f_{12}$ are within the realm of $\gamma$\,Doradus pulsations \cite{Grigahcene+10apj,Handler99mn}, which are found between approximately 0.3 and 3\cd. Several of the detected frequencies are multiples of $f_{\rm orb}$, which suggests that they are tidally perturbed or excited pulsation modes \cite{Guo++17apj}. The last set of oscillations detected ($f_{13}$ to $f_{35}$) have frequencies consistent with $\delta$\,Scuti pulsations \cite{Breger00aspc,Grigahcene+10apj}.

We have therefore detected pulsations in RR~Lyn arising from tidal effects, and from the $\gamma$\,Dor and $\delta$\,Sct mechanisms. Some of them are integer multiples of $f_{\rm orb}$ and some adjacent frequencies in the list are separated by $f_{\rm orb}$. Both stars have physical properties (\Teff\ and \logg) consistent with the $\gamma$\,Dor and $\delta$\,Sct instability strips \cite{Grigahcene+10apj,Henry++07aj,HandlerShobbrook02mn,Uytterhoeven+11aa} so we are not able to assign specific pulsation frequencies to individual stars. Fig.\,\ref{fig:tess} shows a slight increase in pulsation amplitude during primary eclipse, and the opposite during secondary eclipse, implying that the secondary star is the source of most of the pulsations, but our data are not sufficient to allow definitive conclusions.

\begin{table} \centering
\caption{\em Significant pulsation frequencies found in the \tess\ light
curve of RR~Lyn after subtraction of the effects of binarity. Frequencies
that are close to a multiple of the orbital frequency $f_{\rm orb}$ are
labelled with the multiple in the Notes column. \label{tab:freq}}
\setlength{\tabcolsep}{12pt}
\begin{tabular}{lcccc}
{\em Label} & {\em Frequency (d$^{-1}$)} & {\em Amplitude (mmag)} & {\em Phase} & {\em Note} \\[3pt]
$f_{ 1}$ & $ 0.1064 \pm 0.0005$ & $0.250 \pm 0.004$ & $0.912 \pm 0.003$ & $\approx$$f_{\rm orb}$ \\       
$f_{ 2}$ & $ 0.1558 \pm 0.0009$ & $0.164 \pm 0.004$ & $0.249 \pm 0.004$ \\                                
$f_{ 3}$ & $ 0.2071 \pm 0.0003$ & $0.383 \pm 0.004$ & $0.882 \pm 0.002$ & $\approx$2$f_{\rm orb}$ \\      
$f_{ 4}$ & $ 0.2983 \pm 0.0007$ & $0.139 \pm 0.004$ & $0.681 \pm 0.005$ & $\approx$3$f_{\rm orb}$ \\      
$f_{ 5}$ & $ 0.3990 \pm 0.0008$ & $0.096 \pm 0.004$ & $0.565 \pm 0.006$ & $\approx$4$f_{\rm orb}$ \\      
$f_{ 6}$ & $ 0.5016 \pm 0.0007$ & $0.114 \pm 0.004$ & $0.733 \pm 0.006$ & $\approx$5$f_{\rm orb}$ \\      
$f_{ 7}$ & $ 0.5548 \pm 0.0003$ & $0.272 \pm 0.004$ & $0.559 \pm 0.002$ \\                                
$f_{ 8}$ & $ 0.7106 \pm 0.0009$ & $0.095 \pm 0.004$ & $0.348 \pm 0.007$ & $\approx$7$f_{\rm orb}$ \\      
$f_{ 9}$ & $ 0.7884 \pm 0.0007$ & $0.111 \pm 0.004$ & $0.320 \pm 0.005$ \\                                
$f_{10}$ & $ 0.9062 \pm 0.0009$ & $0.089 \pm 0.004$ & $0.829 \pm 0.007$ & $\approx$9$f_{\rm orb}$ \\      
$f_{11}$ & $ 1.6206 \pm 0.0005$ & $0.156 \pm 0.004$ & $0.711 \pm 0.004$ \\                                
$f_{12}$ & $ 1.7137 \pm 0.0009$ & $0.093 \pm 0.004$ & $0.552 \pm 0.006$ & $\approx$17$f_{\rm orb}$ \\     
$f_{13}$ & $ 3.5072 \pm 0.0011$ & $0.071 \pm 0.004$ & $0.467 \pm 0.009$ \\                                
$f_{14}$ & $ 5.0973 \pm 0.0015$ & $0.050 \pm 0.004$ & $0.674 \pm 0.012$ \\                                
$f_{15}$ & $ 5.2227 \pm 0.0013$ & $0.061 \pm 0.004$ & $0.327 \pm 0.010$ \\                                
$f_{16}$ & $ 5.7965 \pm 0.0016$ & $0.050 \pm 0.004$ & $0.807 \pm 0.012$ \\                                
$f_{17}$ & $ 5.4925 \pm 0.0017$ & $0.045 \pm 0.004$ & $0.788 \pm 0.013$ \\                                
$f_{18}$ & $ 5.6198 \pm 0.0012$ & $0.069 \pm 0.004$ & $0.754 \pm 0.009$ \\                                
$f_{19}$ & $ 6.1746 \pm 0.0013$ & $0.056 \pm 0.004$ & $0.362 \pm 0.010$ \\                                
$f_{20}$ & $ 6.2468 \pm 0.0008$ & $0.092 \pm 0.004$ & $0.277 \pm 0.006$ \\                                
$f_{21}$ & $ 6.3988 \pm 0.0007$ & $0.106 \pm 0.004$ & $0.861 \pm 0.005$ \\                                
$f_{22}$ & $ 6.5374 \pm 0.0012$ & $0.067 \pm 0.004$ & $0.189 \pm 0.009$ \\                                
$f_{23}$ & $ 6.5811 \pm 0.0011$ & $0.072 \pm 0.004$ & $0.063 \pm 0.009$ \\                                
$f_{24}$ & $ 6.7768 \pm 0.0008$ & $0.088 \pm 0.004$ & $0.221 \pm 0.006$ \\                                
$f_{25}$ & $ 6.9022 \pm 0.0004$ & $0.186 \pm 0.004$ & $0.906 \pm 0.003$ \\                                
$f_{26}$ & $ 6.9630 \pm 0.0008$ & $0.103 \pm 0.004$ & $0.536 \pm 0.006$ \\                                
$f_{27}$ & $ 7.0827 \pm 0.0007$ & $0.114 \pm 0.004$ & $0.669 \pm 0.005$ \\                                
$f_{28}$ & $ 7.1872 \pm 0.0008$ & $0.102 \pm 0.004$ & $0.273 \pm 0.006$ \\                                
$f_{29}$ & $ 7.3696 \pm 0.0003$ & $0.244 \pm 0.004$ & $0.435 \pm 0.002$ \\                                
$f_{30}$ & $ 7.5615 \pm 0.0014$ & $0.050 \pm 0.004$ & $0.461 \pm 0.011$ \\                                
$f_{31}$ & $ 7.6983 \pm 0.0009$ & $0.086 \pm 0.004$ & $0.704 \pm 0.007$ \\                                
$f_{32}$ & $ 7.8237 \pm 0.0014$ & $0.055 \pm 0.004$ & $0.267 \pm 0.011$ \\                                
$f_{33}$ & $10.8958 \pm 0.0010$ & $0.078 \pm 0.004$ & $0.019 \pm 0.008$ \\                                
$f_{34}$ & $11.1864 \pm 0.0009$ & $0.091 \pm 0.004$ & $0.892 \pm 0.007$ \\                                
$f_{35}$ & $18.8836 \pm 0.0007$ & $0.108 \pm 0.004$ & $0.641 \pm 0.006$ \\[5pt]                           
\end{tabular}
\end{table}

\section*{Summary}

RR~Lyn is a dEB with several interesting features. The primary component is a slightly-evolved 1.9\Msun\ star and shows chemical peculiarities of the Am type. The secondary component is an unevolved 1.5\Msun\ star and may be metal-poor. There may be a third body in the system causing changes in the eclipse times due to the light-time effect. We have used the \tess\ light curve and published RVs to determine the masses and radii of the stars to high precision (0.3\% in mass and 0.8\% in radius). These properties, plus their \Teff\ values, match the predictions of theoretical models for a solar chemical composition and an age in the region of 1~Gyr.

The \tess\ light curve of RR~Lyn shows clear evidence for pulsations. We have measured 35 pulsation frequencies from these data, and find that they are consistent with being tidally perturbed $\gamma$\,Dor and $\delta$\,Sct pulsations. We tentatively assign the pulsations (or at least the majority of them) to star~B, in agreement with past observations that the incidence of pulsations in Am stars is lower than for normal A-stars \cite{Balona+11mn,Smalley+11aa2,Smalley+17mn}. The photospheric abundances of the stars should be carefully measured to investigate the metallic-line nature of star~A and to determine the chemical composition of the system from star~B. When combined with the precisely-known masses, radii and oscillation frequencies, it may be possible to place stringent constraints on the stellar physics incorporated into the current generation of theoretical evolutionary models.


\section*{Acknowledgements}

I am grateful to Timothy Van Reeth, Pierre Maxted, Dariusz Graczyk and the referee for helpful comments on this work.
This paper includes data collected by the \tess\ mission. Funding for the \tess\ mission is provided by the NASA's Science Mission Directorate.
The following resources were used in the course of this work: the NASA Astrophysics Data System; the SIMBAD database operated at CDS, Strasbourg, France; and the ar$\chi$iv scientific paper preprint service operated by Cornell University.




\begin{thebibliography}{10}
\newcommand{\enquote}[1]{`(#1)'}

\bibitem{Andersen++90apj}
J.~{Andersen}, J.~V. {Clausen} \& B.~{Nordstr{\"o}m}, \textit{ApJ},
  \textbf{363}, L33, 1990.

\bibitem{Torres++10aarv}
G.~{Torres}, J.~{Andersen} \& A.~{Gim{\'e}nez}, \textit{A\&ARv}, \textbf{18},
  67, 2010.

\bibitem{ClaretTorres18apj}
A.~{Claret} \& G.~{Torres}, \textit{ApJ}, \textbf{859}, 100, 2018.

\bibitem{Tkachenko+20aa}
A.~{Tkachenko} \textit{et~al.}, \textit{A\&A}, \textbf{637}, A60, 2020.

\bibitem{Aerts++10book}
C.~{Aerts}, J.~{Christensen-Dalsgaard} \& D.~W. {Kurtz},
  \textit{{Asteroseismology}} ({Astron.\ and Astroph.\ Library, Springer
  Netherlands, Amsterdam}), 2010.

\bibitem{Aerts+03sci}
C.~{Aerts} \textit{et~al.}, \textit{Science}, \textbf{300}, 1926, 2003.

\bibitem{Briquet+07mn}
M.~{Briquet} \textit{et~al.}, \textit{MNRAS}, \textbf{381}, 1482, 2007.

\bibitem{Garcia+13aa}
A.~{Garc{\'{\i}}a Hern{\'a}ndez} \textit{et~al.}, \textit{A\&A}, \textbf{559},
  A63, 2013.

\bibitem{Bedding+20nat}
T.~R. {Bedding} \textit{et~al.}, \textit{Nature}, \textbf{581}, 147, 2020.

\bibitem{Me+11mn}
J.~{Southworth} \textit{et~al.}, \textit{MNRAS}, \textbf{414}, 2413, 2011.

\bibitem{Hambleton+13mn}
K.~M. {Hambleton} \textit{et~al.}, \textit{MNRAS}, \textbf{434}, 925, 2013.

\bibitem{Maceroni+14aa}
C.~{Maceroni} \textit{et~al.}, \textit{A\&A}, \textbf{563}, A59, 2014.

\bibitem{Guo+16apj}
Z.~{Guo} \textit{et~al.}, \textit{ApJ}, \textbf{826}, 69, 2016.

\bibitem{Maceroni+13aa}
C.~{Maceroni} \textit{et~al.}, \textit{A\&A}, \textbf{552}, A60, 2013.

\bibitem{Guo+19apj}
Z.~{Guo} \textit{et~al.}, \textit{ApJ}, \textbf{885}, 46, 2019.

\bibitem{Clausen96aa}
J.~V. {Clausen}, \textit{A\&A}, \textbf{308}, 151, 1996.

\bibitem{Me+20mn}
J.~{Southworth} \textit{et~al.}, \textit{MNRAS}, \textbf{497}, L19, 2020.

\bibitem{LeeHong21aj}
J.~W. {Lee} \& K.~{Hong}, \textit{AJ}, \textbf{161}, 32, 2021.

\bibitem{Me++21mn}
J.~{Southworth}, D.~M. {Bowman} \& K.~{Pavlovski}, \textit{MNRAS},
  \textbf{501}, L65, 2021.

\bibitem{Gaulme+16apj}
P.~{Gaulme} \textit{et~al.}, \textit{ApJ}, \textbf{832}, 121, 2016.

\bibitem{Themessl+18mn}
N.~{Theme{\ss}l} \textit{et~al.}, \textit{MNRAS}, \textbf{478}, 4669, 2018.

\bibitem{Benbakoura+21aa}
M.~{Benbakoura} \textit{et~al.}, \textit{A\&A}, \textbf{648}, A113, 2021.

\bibitem{Maceroni+09aa}
C.~{Maceroni} \textit{et~al.}, \textit{A\&A}, \textbf{508}, 1375, 2009.

\bibitem{Fuller17mn}
J.~{Fuller}, \textit{MNRAS}, \textbf{472}, 1538, 2017.

\bibitem{Bowman+19apj}
D.~M. {Bowman} \textit{et~al.}, \textit{ApJ}, \textbf{883}, L26, 2019.

\bibitem{Fuller+20mn}
J.~{Fuller} \textit{et~al.}, \textit{MNRAS}, \textbf{498}, 5730, 2020.

\bibitem{CampbellWright00apj}
W.~W. {Campbell} \& W.~H. {Wright}, \textit{ApJ}, \textbf{12}, 254, 1900.

\bibitem{Baglin+73aa}
A.~{Baglin} \textit{et~al.}, \textit{A\&A}, \textbf{23}, 221, 1973.

\bibitem{Breger00aspc}
M.~{Breger}, in \textit{Delta Scuti and Related Stars} ({M.~Breger \&
  M.~Montgomery}, ed.), 2000, \textit{Astronomical Society of the Pacific
  Conference Series}, vol. 210, pp. 3--42.

\bibitem{Murphy+19mn}
S.~J. {Murphy} \textit{et~al.}, \textit{MNRAS}, \textbf{485}, 2380, 2019.

\bibitem{Grigahcene+10apj}
A.~{Grigahc{\`e}ne} \textit{et~al.}, \textit{ApJ}, \textbf{713}, L192, 2010.

\bibitem{Garcia+15apj}
A.~{Garc{\'{\i}}a Hern{\'a}ndez} \textit{et~al.}, \textit{ApJ}, \textbf{811},
  L29, 2015.

\bibitem{Kaye+99pasp}
A.~B. {Kaye} \textit{et~al.}, \textit{PASP}, \textbf{111}, 840, 1999.

\bibitem{Henry++07aj}
G.~W. {Henry}, F.~C. {Fekel} \& S.~M. {Henry}, \textit{AJ}, \textbf{133}, 1421,
  2007.

\bibitem{Balona++15mn}
L.~A. {Balona}, J.~{Daszy{\'n}ska-Daszkiewicz} \& A.~A. {Pamyatnykh},
  \textit{MNRAS}, \textbf{452}, 3073, 2015.

\bibitem{Me15debcat}
J.~{Southworth}, in \textit{Living Together: Planets, Host Stars and Binaries}
  (S.~M. {Rucinski}, G.~{Torres} \& M.~{Zejda}, eds.), 2015,
  \textit{Astronomical Society of the Pacific Conference Series}, vol. 496, p.
  321.

\bibitem{Me20obs}
J.~{Southworth}, \textit{The Observatory}, \textbf{140}, 247, 2020.

\bibitem{HoffleitJaschek91}
D.~{Hoffleit} \& C.~. {Jaschek}, \textit{{The Bright Star Catalogue}} (New
  Haven, Conn.: Yale University Observatory, 1991, 5th ed.), 1991.

\bibitem{CannonPickering18anhar2}
A.~J. {Cannon} \& E.~C. {Pickering}, \textit{Annals of Harvard College
  Observatory}, \textbf{92}, 1, 1918.

\bibitem{Gaia21aa}
{Gaia Collaboration} \textit{et~al.}, \textit{A\&A}, \textbf{649}, A1, 2021.

\bibitem{Stassun+19aj}
K.~G. {Stassun} \textit{et~al.}, \textit{AJ}, \textbf{158}, 138, 2019.

\bibitem{Hog+00aa}
E.~{H{\o}g} \textit{et~al.}, \textit{A\&A}, \textbf{355}, L27, 2000.

\bibitem{Cutri+03book}
R.~M. {Cutri} \textit{et~al.}, \textit{{2MASS All Sky Catalogue of Point
  Sources}} (The IRSA 2MASS All-Sky Point Source Catalogue, NASA/IPAC Infrared
  Science Archive, Caltech, US), 2003.

\bibitem{LevatoAbt78pasp}
H.~{Levato} \& H.~A. {Abt}, \textit{PASP}, \textbf{90}, 429, 1978.

\bibitem{Khaliullin++01arep}
K.~F. {Khaliullin}, A.~I. {Khaliullina} \& A.~V. {Krylov}, \textit{Astronomy
  Reports}, \textbf{45}, 888, 2001.

\bibitem{Adams12apj}
W.~S. {Adams}, \textit{ApJ}, \textbf{35}, 163, 1912.

\bibitem{Harper15pdo}
W.~E. {Harper}, \textit{Publications of the Dominion Observatory Ottawa},
  \textbf{2}, 165, 1915.

\bibitem{DouglasPopper63pasp}
B.~C. {Douglas} \& D.~M. {Popper}, \textit{PASP}, \textbf{75}, 411, 1963.

\bibitem{Popper71apj}
D.~M. {Popper}, \textit{ApJ}, \textbf{169}, 549, 1971.

\bibitem{Kondo76antok}
M.~{Kondo}, \textit{Annals of the Tokyo Astronomical Observatory}, \textbf{16},
  1, 1976.

\bibitem{TomkinFekel06aj}
J.~{Tomkin} \& F.~C. {Fekel}, \textit{AJ}, \textbf{131}, 2652, 2006.

\bibitem{Bensch+14ibvs}
K.~{Bensch} \textit{et~al.}, \textit{IBVS}, \textbf{6121}, 1, 2014.

\bibitem{Huffer31paas}
C.~M. {Huffer}, in \textit{Publications of the American Astronomical Society},
  1931, vol.~6, p. 365.

\bibitem{MagalashviliKumsishvili59abaob}
N.~L. {Magalashvili} \& J.~I. {Kumsishvili}, \textit{Abastumanskaia
  Astrofizicheskaia Observatoriia Byulleten}, \textbf{24}, 13, 1959.

\bibitem{Botsula60baoe}
R.~A. {Botsula}, \textit{Bull.\ Astron.\ Obs.\ Engelgardt}, \textbf{35}, 43,
  1960.

\bibitem{Linnell66aj}
A.~P. {Linnell}, \textit{AJ}, \textbf{71}, 458, 1966.

\bibitem{Lavrov++88trkaz}
M.~I. {Lavrov}, N.~V. {Lavrova} \& Y.~F. {Shabalov}, \textit{Trudy Kazanskaia
  Gorodkoj Astronomicheskoj Observatorii}, \textbf{51}, 19, 1988.

\bibitem{KhaliullinKhaliullina02arep}
K.~F. {Khaliullin} \& A.~I. {Khaliullina}, \textit{Astronomy Reports},
  \textbf{46}, 119, 2002.

\bibitem{Me21obs4}
J.~{Southworth}, \textit{The Observatory}, \textbf{141}, 190, 2021.

\bibitem{Budding74apss}
E.~{Budding}, \textit{Ap\&SS}, \textbf{30}, 433, 1974.

\bibitem{Botsula68sovast}
R.~A. {Botsula}, \textit{Soviet Astronomy}, \textbf{11}, 1000, 1968.

\bibitem{Roman49apj}
N.~G. {Roman}, \textit{ApJ}, \textbf{110}, 205, 1949.

\bibitem{Cowley+69aj}
A.~{Cowley} \textit{et~al.}, \textit{AJ}, \textbf{74}, 375, 1969.

\bibitem{AbtBidelman69apj}
H.~A. {Abt} \& W.~P. {Bidelman}, \textit{ApJ}, \textbf{158}, 1091, 1969.

\bibitem{TitusMorgan40apj}
J.~{Titus} \& W.~W. {Morgan}, \textit{ApJ}, \textbf{92}, 256, 1940.

\bibitem{Conti70pasp}
P.~S. {Conti}, \textit{PASP}, \textbf{82}, 781, 1970.

\bibitem{AbtMorrell95apjs}
H.~A. {Abt} \& N.~I. {Morrell}, \textit{ApJS}, \textbf{99}, 135, 1995.

\bibitem{Ricker+15jatis}
G.~R. {Ricker} \textit{et~al.}, \textit{Journal of Astronomical Telescopes,
  Instruments, and Systems}, \textbf{1}, 014003, 2015.

\bibitem{Jenkins+16spie}
J.~M. {Jenkins} \textit{et~al.}, in \textit{Proc.\ SPIE}, 2016, \textit{Society
  of Photo-Optical Instrumentation Engineers (SPIE) Conference Series}, vol.
  9913, p. 99133E.

\bibitem{Me++04mn2}
J.~{Southworth}, P.~F.~L. {Maxted} \& B.~{Smalley}, \textit{MNRAS},
  \textbf{351}, 1277, 2004.

\bibitem{Me13aa}
J.~{Southworth}, \textit{A\&A}, \textbf{557}, A119, 2013.

\bibitem{Maxted+20mn}
P.~F.~L. {Maxted} \textit{et~al.}, \textit{MNRAS}, \textbf{498}, 332, 2020.

\bibitem{Kopal50}
Z.~{Kopal}, \textit{Harvard College Observatory Circular}, \textbf{454}, 1,
  1950.

\bibitem{Claret18aa}
A.~{Claret}, \textit{A\&A}, \textbf{618}, A20, 2018.

\bibitem{Me++04mn}
J.~{Southworth}, P.~F.~L. {Maxted} \& B.~{Smalley}, \textit{MNRAS},
  \textbf{349}, 547, 2004.

\bibitem{Me08mn}
J.~{Southworth}, \textit{MNRAS}, \textbf{386}, 1644, 2008.

\bibitem{Me21obs5}
J.~{Southworth}, \textit{The Observatory, in press, \texttt{arXiv:2106.04323}},
  2021.

\bibitem{Prsa+16aj}
A.~{Pr{\v s}a} \textit{et~al.}, \textit{AJ}, \textbf{152}, 41, 2016.

\bibitem{Hilditch01book}
R.~W. {Hilditch}, \textit{{An Introduction to Close Binary Stars}} (Cambridge
  University Press, Cambridge, UK), 2001.

\bibitem{Me++05aa}
J.~{Southworth}, P.~F.~L. {Maxted} \& B.~{Smalley}, \textit{A\&A},
  \textbf{429}, 645, 2005.

\bibitem{Popper80araa}
D.~M. {Popper}, \textit{ARA\&A}, \textbf{18}, 115, 1980.

\bibitem{Lallement+14aa}
R.~{Lallement} \textit{et~al.}, \textit{A\&A}, \textbf{561}, A91, 2014.

\bibitem{Lallement+18aa}
R.~{Lallement} \textit{et~al.}, \textit{A\&A}, \textbf{616}, A132, 2018.

\bibitem{PecautMamajek13apjs}
M.~J. {Pecaut} \& E.~E. {Mamajek}, \textit{ApJS}, \textbf{208}, 9, 2013.

\bibitem{Graczyk+19apj}
D.~{Graczyk} \textit{et~al.}, \textit{ApJ}, \textbf{872}, 85, 2019.

\bibitem{Bressan+12mn}
A.~{Bressan} \textit{et~al.}, \textit{MNRAS}, \textbf{427}, 127, 2012.

\bibitem{LenzBreger05coast}
P.~{Lenz} \& M.~{Breger}, \textit{Communications in Asteroseismology},
  \textbf{146}, 53, 2005.

\bibitem{Breger+93aa}
M.~{Breger} \textit{et~al.}, \textit{A\&A}, \textbf{271}, 482, 1993.

\bibitem{Kuschnig+97aa}
R.~{Kuschnig} \textit{et~al.}, \textit{A\&A}, \textbf{328}, 544, 1997.

\bibitem{LoumosDeeming78apss}
G.~L. {Loumos} \& T.~J. {Deeming}, \textit{Ap\&SS}, \textbf{56}, 285, 1978.

\bibitem{Handler99mn}
G.~{Handler}, \textit{MNRAS}, \textbf{309}, L19, 1999.

\bibitem{Guo++17apj}
Z.~{Guo}, D.~R. {Gies} \& J.~{Fuller}, \textit{ApJ}, \textbf{834}, 59, 2017.

\bibitem{HandlerShobbrook02mn}
G.~{Handler} \& R.~R. {Shobbrook}, \textit{MNRAS}, \textbf{333}, 251, 2002.

\bibitem{Uytterhoeven+11aa}
K.~{Uytterhoeven} \textit{et~al.}, \textit{A\&A}, \textbf{534}, A125, 2011.

\bibitem{Balona+11mn}
L.~A. {Balona} \textit{et~al.}, \textit{MNRAS}, \textbf{414}, 792, 2011.

\bibitem{Smalley+11aa2}
B.~{Smalley} \textit{et~al.}, \textit{A\&A}, \textbf{535}, A3, 2011.

\bibitem{Smalley+17mn}
B.~{Smalley} \textit{et~al.}, \textit{MNRAS}, \textbf{465}, 2662, 2017.

\end{thebibliography}

\end{document}